\documentclass[aps, prd, twocolumn, lengthcheck, superscriptaddress, showpacs, letterpaper, nofootinbib]{revtex4-1}

\usepackage{epsfig}
\usepackage[usenames]{color}
\usepackage{graphicx}
\usepackage{amsmath}
\usepackage{epstopdf}

\newcommand\sect[1]{\emph{#1}---}
\def\bi{\bibitem}

\def\la{\langle}\def\ra{\rangle}
\def\be{\begin{eqnarray}}\def\ee{\end{eqnarray}}
\def\lsim{\mathrel{\rlap{\lower3pt\hbox{\hskip1pt$\sim$}}
     \raise1pt\hbox{$<$}}} 
\def\gsim{\mathrel{\rlap{\lower3pt\hbox{\hskip1pt$\sim$}}
     \raise1pt\hbox{$>$}}} 
\def\del{\partial}

\allowdisplaybreaks

\def\la{\langle}\def\ra{\rangle}
\def\be{\begin{eqnarray}}\def\ee{\end{eqnarray}}
\def\lsim{\mathrel{\rlap{\lower3pt\hbox{\hskip1pt$\sim$}}
     \raise1pt\hbox{$<$}}} 
\def\gsim{\mathrel{\rlap{\lower3pt\hbox{\hskip1pt$\sim$}}
     \raise1pt\hbox{$>$}}} 
\def\le{ \begin{array}{ll}}\def\re{\end{array}}

\def\lear{ \left( \begin{array}{cc}}\def\rear{\end{array} \right)}

\def\le{ \left( \begin{array}{cc}}\def\re{\end{array} \right)}

\def\bi{\bibitem}

\def\eft-hls{{\it EFT}$_{\rm bsHLS}$}
\def\skyrmion-hls{{\it Skyrmion}$_{\rm sHLS}$}

\begin{document}

\title{A pseudo-conformal equation of state in compact-star matter \\ from topology change and hidden symmetries of QCD}
\author{Yong-Liang Ma}
\affiliation{Center for Theoretical Physics and College of Physics, Jilin University, Changchun, 130012, China}
\author{Hyun Kyu Lee}
\affiliation{Department of Physics, Hanyang University, Seoul 133-791, Korea}
\author{Won-Gi Paeng}
\affiliation{Rare Isotope Science Project, Institute for Basic Science, Daejeon 305-811, Korea}
\author{Mannque Rho}
\affiliation{Institut de Physique Th\'eorique,
	CEA Saclay, 91191 Gif-sur-Yvette c\'edex, France }

\date{\today}

\begin{abstract}
We construct a new effective field theory approach to the equation of state (EoS), dubbed pseudo-confomal model ``PCM," for nuclear and compact star matter entirely in terms of effective hadron degrees of freedom. The possible transition at $n\sim (2-4) n_0$ (where $n_0$ is the normal nuclear matter density) from hadron degrees of freedom to strongly-coupled quark degrees of freedom,  giving rise to  a soft-to-hard changeover in the EoS  that can accommodate the massive stars observed, is effectuated by the topology change at $n_{1/2}\gsim 2n_0$ from skyrmions to half-skyrmions without involving local order-parameter fields. The mechanism exploits possible emergence of hidden scale and local symmetries of QCD at high density,  leading to a precocious ``pseudo-conformal" sound velocity $v_s^2=1/3$ (in unit of $c=1$) for $n\gsim 3n_0$. The resulting prediction signals a drastic departure from  standard nuclear many-body theory in the density regime involved in the massive stars. We suggest that  the tidal deformability  implemented  in gravitational waves coming from coalescing neutron stars in LIGO/Virgo-type observations could pin down the location of the topology change density $n_{1/2}$.

\end{abstract}

\maketitle

\sect{\bf Introduction}
It has been found in a recent effort to go in one unique effective field theory (EFT) from the well-studied density regime of normal equilibrium nuclear matter $n_0\simeq 0.16$ fm$^{-3}$ to the more or less uncharted density regime of compact-star matter $\sim (5-7)n_0$,  that topology and certain symmetries in the strong interactions,  both invisible in QCD in the matter-free vacuum, could emerge in dense medium and play powerful roles~\cite{PKLMR}. They enter in accounting for the possible -- and expected -- change of relevant degrees of freedom from hadrons to quarks/gluons as density goes above some high density, denoted $n_{1/2} > 2n_0$.

At low energy -- and low density --  topology, figuring in the baryon structure in the  large number-of-colors limit ($N_c\to \infty$), is ``hidden"  in QCD. The correspondence between the two, topology and QCD,  could be phrased in terms of the ``Cheshire Cat" principle formulated  in 1980's~\cite{cheshirecat}. In this formulation, the QCD degrees of freedom, e.g., the MIT bag,  can be smoothly traded in for a topological soliton, i.e., the skyrmion, so as to represent  hadron-quark continuity.  As density increases above $n_{1/2}$,  the possible hadrons-to-quarks/gluons transition is then to be translated into a topology change. In this smooth cross-over,  two symmetries invisible in QCD in the matter-free space,   one, hidden local (gauge) symmetry  and the other,  hidden scale symmetry, are argued to intervene.

In this Letter, we take the effective field theory  framework  constructed with  those hidden ingredients suitably incorporated in a description that involves only hadronic variables to encompass the whole range from normal density to high density in nuclear dynamics,  and   make certain predictions for compact-star matter that have not been made in  ``standard nuclear EFTs"\footnote{By standard nuclear EFT, we mean all those chiral effective field theory approaches anchored on chiral Lagrangian (involving pions and nucleons only)  and also EDF (energy density functional) approaches including relativistic mean field treatments.}. These predictions, if confirmed, would reveal the emergence of  hidden symmetries at a precocious density $n\gsim 3n_0$.

\sect{\bf Hidden symmetries}
Hidden local symmetry (HLS) we are concerned with is associated with the low-lying vector mesons, particularly  the $\rho$. In principle, one can include the infinite towers of vector mesons as in holographic QCD. We will assume, without losing predictive power, that we can have the high towers integrated out. The lowest lying $\omega$ meson can be treated in the same way as the $\rho$ at low density, but there is a good reason to believe that the flavor $U(2)$ symmetry is badly broken at density $n\gsim n_{1/2}$. QCD of course has no flavor local symmetry, so HLS is absent in the vacuum. However  it is gauge-equivalent to nonlinear sigma model at low energy which governs low-energy hadron dynamics (e.g., chiral Lagrangian) and can emerge dynamically~\cite{HY:PR,yamawaki2018}, with the local gauge symmetry manifesting explicitly at what is called ``vector manifestation (VM) fixed point" at which the vector meson $\rho$  becomes massless~\cite{HY:PR}.

As for scale invariance, while it is present in the QCD Lagrangian in the absence of quark-mass terms, it is  broken by the quantum effect, i.e., the trace anomaly.  In the vacuum, scale symmetry is therefore not visible. However high density could change the situation. It has been suggested that it can appear induced by strong nuclear correlations~\cite{PKLMR}. One can say that  this is natural because this symmetry is actually there but hidden in QCD~\cite{CCT,yamawaki2018}. The way to see this is to use  linear sigma model or equivalently NJL model~\cite{yamawaki2018}. There is an intrinsic parameter in the model that can be dialed externally, say, by temperature or density.  By dialing the parameter, one could induce a continuous transition from a state described by nonlinear sigma model to one given by a nearly scale-invariant state. It was suggested ~\cite{PKLMR} that the dialing can be effectuated in nuclear interactions by density.

It was found in \cite{PKLMR,LMR-gA} that with the properties described above suitably incorporated in a nuclear effective field theory Lagrangian, one could describe fairly well both low and high density equations of state. The Lagrangian, denoted $bs$HLS ($b$ standing for baryons, $s$ for scalar field $\sigma$ and HLS for hidden local fields), consists, in addition to the degrees of freedom present in the standard chiral effective field theory (S$\chi$EFT for short), i.e., the pseudo-Nambu-Goldstone (pNG) bosons $\pi$ and the nucleons, of the vector mesons brought in as hidden local fields and the scalar $\sigma$ as a pNG for spontaneously broken scale symmetry. This Lagrangian with the ``bare" parameters, suitably matched via current correlators to QCD at a matching scale $\Lambda^\ast$ appropriate for nuclear physics, and the topology change at $n_{1/2}$ implemented, when treated in the  Wilsonian renormalization-group approach using the $V_{lowk}$ ($V_{lowk}$RG for short)~\cite{PKLMR}, is found to give (a) at low density near $n_0$ a surprisingly good description of certain low-momentum transfer properties such as the famous -- and long-standing -- ``quenched" $g_A$ problem in the Gamow-Teller transitions in nuclei~\cite{LMR-gA} as well as  the bulk properties of nuclear matter~\cite{PKLMR}, and (b) at compact-star density, the equation of state for massive neutron stars~\cite{PKLMR}.

Now the principal point of this paper is that  the Lagrangian $bs$HLS, so constructed, leads -- with slight refinements -- to some strikingly new predictions on hitherto undiscovered  structure of compact stars  that have not been found in  S$\chi$EFT in the literature. We  suggest that those predictions represent a precursor signal for a precocious manifestation of the hidden topology and symmetries. Here there is an uncanny analogy to what's currently taking place in condensed matter physics. What is also quite appealing is that the density at which the topology change, specified below, takes place could be pinned down once the tidal deformability seen in recent LIGO/Virgo observations in coalescing neutron stars~\cite{Ligo/Virgo}  can be precisely determined.

\sect{\bf Topology change and the nuclear tensor force}
Among the key ingredients that figure in giving the results of \cite{PKLMR,LMR-gA} is the observation that when baryonic matter is described in terms of skyrmions, there is a topology change at a density $n_{1/2}\gsim 2.0n_0$ involving skyrmions fractionalizing into half-skyrmions. We interpret this  as having the strong-coupling quark degrees of freedom as exploited in, say, \cite{baymetal} in the density range $\sim (2-4)n_0$ traded in for topology as a Cheshire-Cat phenomenon.

The topology change brings out several important effects which can be summarized in two main observations~\cite{PKLMR} that constitute the basis for the predictions to be given below.

$\bullet$ {\bf A}: The first observation, the most crucial of all, is that there is a cusp at $n_{1/2}$ in the symmetry energy $E_{sym}$ in the EoS of compact-star matter. This appears when the skyrmion neutron matter is collective-quantized~\cite{cusp}. It is of leading order in $N_c$, hence topological and robust. This cusp structure can be incorporated (or translated) into the effective Lagrangian  $bs$HLS~\cite{PKLMR}. This is done by matching at $\Lambda^\ast$ the bare parameters of $bs$HLS to QCD via suitable correlators as mentioned above. {This matching could be most effectively done at the chiral scale $\Lambda_\chi\sim 1$ GeV but in practice in nuclear physics, it is typically of the vector meson mass $\sim m_\rho$.}  Inside nuclear medium, the Lagrangian is made to track the vacuum changes  with the ``intrinsic density dependence (IDD)" encoded in the QCD condensates that are brought in by matching  to QCD. The cusp is manifested in  the symmetry energy when the Lagrangian $bs$HLS is treated at the mean field level, which corresponds to the semiclassical approximation. Going beyond the mean field in the $V_{lowk}$RG smoothens the cusp. The cusp exposes the intricate structure of the nuclear tensor force -- with the intrinsic density effect -- which decreases as density goes toward $n_{1/2}$ and then increases past $n_{1/2}$.  Here the  vector manifestation (VM) of the $\rho$ meson  is found to play an indispensable role in producing the cusp. It is worth pointing out that the decreasing strength of the tensor force  near $n_0$ going toward $n_{1/2}$ from below is confirmed  beautifully in the long life-time of the C14 beta decay~\cite{holt}\footnote{A similar effect is gotten by explicit short-ranged three-body forces. It is essentially the same mechanism as the intrinsic density effect in the two-body tensor force.}.  What will turn out to be noteworthy below is that this cusp, suitably smoothed by corrections to the  semiclassical approximations,  could play an important role  in the tidal deformability in gravitational waves.
\vskip .2cm
$\bullet$ {\bf B}:  Another observation, equally important, that is connected with the topology change, together with emerging symmetries in dense medium, is that the sound velocity in compact stars goes precociously to the conformal velocity $v_s^2=1/3$ at densities $n\gsim 3n_0$,  an observation {\it totally foreign} to S$\chi$EFT. This behavior is due to the fact that  the topology change induces the  parity doubling in the nucleon structure, with the effective in-medium nucleon mass  $m_N^\ast$ going as the dilaton condensate~\cite{PLRS-interplay}  $\la\chi\ra^\ast \propto m_0$ where
$\chi$ is the conformal compensator field and $m_0$ is the chirally invariant mass figuring in the parity-doubled linear sigma model~\cite{detar-kunihiro}.\footnote{ See \cite{protonmass} where the chiral invariant mass $m_0$ is linked to the origin of the proton mass} As a consequence, the trace of the energy momentum tensor (TEMT) $\theta_\mu^\mu$ in the density region $n\gsim n_{1/2}$ is given, both in the mean-field approximation and in the $V_{lowk}$RG,   entirely as a function of the dilaton condensate $\la\chi\ra^\ast$.   That $\del\la\theta^\mu_\mu\ra/\del n=0$ leads to the sound velocity of the star $v_s^2=1/3$, usually associated with conformal sound velocity in the chiral limit. {  One might wonder whether the appearance of $v_s^2=1/3$ so precociously  at non-asymptotic density is not  in  tension with current  nuclear physics constraint and  observation of 2-solar mass neutron stars.  In fact it has been suggested that ``if the conformal limit was found to hold at all densities
this would imply that nuclear physics models breakdown below $2n_0$"~\cite{tews}.  What we find in this work -- which is fully consistent with the constraints mentioned above -- points to that the topology change effectively captures the non-hadronic degrees of freedom needed for the onset of conformal sound speed.  We call the resulting object ``pseudo-conformal sound velocity" because the trace of the energy momentum tensor is nonzero in the region $n\gsim n_{1/2}$ and the symmetry involved is an ``emergent" one~\cite{LMR-gA}.}

 That the in-medium vacuum expectation value (VeV$^\ast$) of the TEMT $\la\theta_\mu^\mu\ra^\ast$ in the mean field is a non-zero constant of density is easy to understand since the mean field of $bs$HLS corresponds to that of Fermi-liquid fixed point. What is important is that the VeV$^\ast$ of the TEMT can be shown to be  a constant also in the full $V_{lowk}$RG calculation which goes beyond the mean field~\cite{PKLMR}.  This means higher-order correlations beyond the mean field do not modify the pseudo-conformal structure in dense medium. This is a crucially important ingredient in developing the pseudo-conformal model given below.

This pseudo-conformal structure was also found to emerge  in the skyrmion crystal simulation as a signal for scale symmetry hidden in QCD~\cite{CCT}  in the half-skyrmion phase. This can be seen in  Fig.~11 of \cite{PKLMR}.

\sect{\bf The pseudo-conformal model}
We now rephrase the two basic observations $\bf A$ and $\bf B$  listed above in a form that encapsulates them in a simple EoS. We do this by replacing what is found in  the $V_{lowk}$RG in the density regime $n\gsim n_{1/2}$  in a manifestly pseudo-conformal form. This approach can be considered as belonging to the class of what is referred to as ``Energy Density Functional (EDF)" popular in nuclear theory community, in particular in the form of relativistic mean field theory. What makes our approach however distinctly different from the usual EDF approaches found in the literature is the intrinsic density dependence  inherited from QCD and the impact of the topology change.

The arguments we develop below rely on the ``master formula" for
 the behavior of  hadron masses in nuclear medium established in \cite{PKLMR}, $\frac{m^\ast_N}{m_N} \approx \frac{m^\ast_\sigma}{m_\sigma} \approx \frac{g_V}{g_V^\ast} \frac{m^\ast_V}{m_V} \approx \frac{f^\ast_\pi}{f_\pi} \approx \frac{\langle \chi \rangle^\ast}{\langle \chi \rangle}\equiv \Phi$ (where $V=(\rho,\omega)$),
 that we assume holds up to the central density of massive stars, i.e., $n_{\rm cen}\approx (5-7)n_0$.
The asterisk stands for the intrinsic density dependence inherited from QCD at the matching scale and the quantities without asterisk are free-space quantities. We denote the two regions divided at $n_{1/2}$ as R(egion)-I for $n< n_{1/2}$ and R(egion{-II for $n\geq n_{1/2}$.

The master scaling relation given above dictates the density dependence of the bare parameters of the effective Lagrangian $bs$HLS, with which the $V_{lowk}$RG is to be performed.

First the region R-I:

Up to $n_0$, there is only one parameter $\Phi$. This is because at low density  $g_V^\ast/g_V=1$. This follows from that the vector meson mass accurately satisfies the KSRF low-energy theorem $m_V^\ast=2 f_\pi^\ast g_V^\ast$ and in the $bs$HLS Lagrangian the vector meson mass scales with $\la\chi\ra^\ast$.  In \cite{PKLMR}, it was shown that the $V_{lowk}$RG with the density scaling of $f_\pi^\ast$ or equivalently $\Phi$  inferred from deeply bound pionic systems describes very well the bulk properties of normal nuclear matter at $n_0$.

We assume that this one parameter EoS can be extrapolated via the density dependence of $\Phi$  up to the  higher topology change density $n_{1/2} > n_0$.

We now turn to the region R-II.

First we would like to capture the observations made in the $V_{lowk}$ in R-II in an EoS that is manifestly pseudo-conformal, that is, possessing a non-vanishing trace of energy-momentum tensor which is independent of density, and exactly reproduces the $V_{lowk}$RG result. In \cite{PKLMR}, such an EoS has been found in the extremely simple form of the  energy per nucleon (minus the vacuum rest mass of the matter written) of the forrm
\begin{equation}
\frac{E}{A} = X^\alpha \left( \frac{n}{n_0} \right)^{1/3} + Y^\alpha \left( \frac{n}{n_0} \right)^{-1} -939\, {\rm MeV} \label{RII}
\end{equation}
with the  coefficients $X^\alpha$ and $Y^\alpha$ for $\alpha=(n_n-n_p)/(n_n+n_p)=(0,1)$ (where $n_{n,p}$ are the neutron and proton densities).  It is easy to verify that for any values of $X$ and $Y$,
\be
\frac{\partial}{\partial n} \la\theta_\mu^\mu\ra=\frac{\partial \epsilon (n)}{\partial n} (1-3{v_s^2})=0\label{derivTEMT}
\ee
with $v_s^2=\frac{\partial P(n)}{\partial n}/\frac{\partial \epsilon(n)}{\partial n}$.  Since $\frac{\partial \epsilon (n)}{\partial n} \neq 0$ in the range of densities involved, we immediately obtain the sound velocity $v_s^2 = 1/3$.

Now we compute the EoS in R-II using (\ref{RII}) with the coefficients determined by the continuity at $n=n_{1/2}$ between R-I given by \cite{PKLMR} and R-II, (\ref{RII}),  for the chemical potential $\mu$ and the pressure $P$
\be
\mu_I=\mu_{II},\ \ P_I=P_{II} \ \ {\rm at} \ \ n=n_{1/2}
\ee
where the subscripts $I$ and $II$ represent the regions R-I and R-II respectively. The union of the EoSs of R-I and R-II so computed  --  that we call ``pseudo-conformal model (PCM)"  -- reproduces  {\it precisely} the results of the $V_{lowk}$RG for massive stars.  This verifies that the EoS from (\ref{RII}) exactly captures the physics of the $V_{lowk}$RG in R-II for $n_{1/2}=2n_0$. We interpret this as saying that the pseudo-conformality given in the extremely simple EoS  Eq. (\ref{RII}) embodies  in R-II the physics of the full $V_{lowk}$RG that results from an intricate interplay of the observables {\bf A} and {\bf B}..

Now we would like to look at the EoS for a system that is built with the topology change at a density higher than $2n_0$, i.e., $n_{1/2} > 2n_0$.  This has to do with the information on the tidal deformability inferred from the recent observation of gravitational waves coming from coalescing neutron stars~\cite{Ligo/Virgo}. It is found that the dimensionless tidal deformability $\Lambda$ -- as defined by the GW170817 ~\cite{Ligo/Virgo} -- calculated in the  $V_{lowk}$RG~\cite{PKLMR} with $n_{1/2}\simeq 2.0n_0$ is found to be $\sim 790$, close to the reported upper bound 800 for $\sim 1.4M_\odot$~\cite{Ligo/Virgo}. Furthermore the more recent analysis of the GW170817 data indicates that the bound  could be tightened to a lower value, with the presently quoted range being between 720 and 170~\cite{abbottetal}. This seems to require that  since $\Lambda$ is sensitive to the symmetry energy in R-I and with the cusp structure entailing the lowering of the symmetry energy going toward $n_{1/2}$ from below,  the topology change density  $n_{1/2}$ be increased to above  $2.0n_0$ so as to bring $\Lambda$ to a lower value.  This seems feasible  since the global properties of  massive stars given within the model are verified to differ very little between $1.5 n_0$ and $2.0n_0$~\cite{dongetal,PKLMR}. Thus we expect  one should be able  to bring $n_{1/2}$ higher than $2.0n_0$,  keeping  unaffected the good overall star properties obtained for $n_{1/2}=2n_0$ in \cite{PKLMR}.

To do this, we assume that the PCM as defined above can be applied to a system with $n_{1/2} > 2n_0$ without affecting the standard star properties seriously. As a first trial, we pick $n_{1/2}=2.6n_0$ and explore what transpires with this value.

\sect{\bf Massive star properties}
First we look at the standard star properties given by the PCM with $n_{1/2}=2.6n_0$. By construction, the sound velocity converges to $v_s^2=1/3$ at $n_{1/2}\gsim 3n_0$ as in the case of $n_{1/2}=2.0n_0$.


The crucial question is: Does this PCM EoS for $n_{1/2}=2.6n_0$ give  correct global star properties as it did for $n_{1/2}=2.0n_0$ reproducing exactly the $V_{lowk}$RG?

The EoSs for both $\alpha=1$ and $0$ are found to come out quite close to the $V_{lowk}$RG results for $n_{1/2}=2n_0$. They are all consistent with currently available experimental bounds.
The particularly important quantity for our discussion is the symmetry energy $E_{sym}$,  which is plotted in Fig.~\ref{EA}.  It comes out consistent with the available experimental bounds, which are at present too broad to be stringent. There is a constraint based on combined data from neutron stars and gravitational waves given in \cite{bal}, $E_{sym} (2n_0)= 46.9\pm 10.1$ MeV, with which the predicted value $E_{sym}^{pred} (2n_0)\simeq 49$ MeV is consistent. However the density probed there is in R-I, so it is not directly relevant to the pseudo-conformal structure.  What  is  remarkable is that the cross-over from soft to hard in $E_{sym}$ -- which mimics  the cusp in the skyrmion description-- resembles closely that of the {\it full} $V_{lowk}$ RG result~\cite{PKLMR}.
\begin{figure}[h]
\begin{center}
\includegraphics[width=5.5cm]{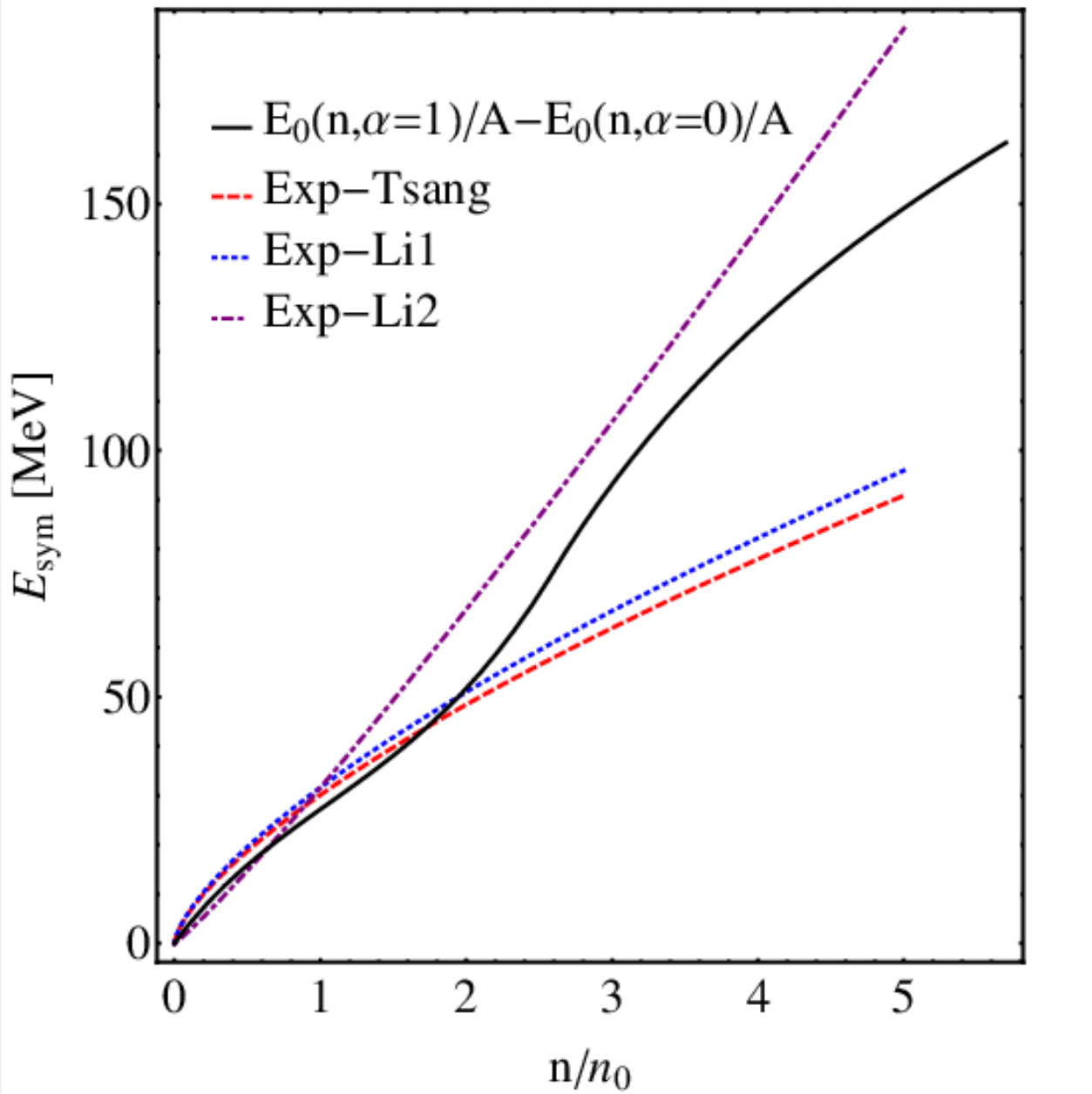}
\caption{$E_{sym}$ (vs density $n$) predicted by the PCM (solid line), which is consistent with the (heavy-ion) experimental bounds quoted from B.A. Li and L.W. Chen, Phy. Rev. C{\bf 72}, 064611 (2005) and M.B. Tsang {\it et al}, Int. J. Mod. Phys. E {\bf 19}, 1631 (2010). }\label{EA}
\end{center}
\end{figure}

In Fig.~\ref{MvsR} are given M vs. R and the central density of the compact star predicted by the PCM with $n_{1/2}=2.6$. They are shown to indicate that the overall structure of the star is insensitive to the density $n_{1/2}$. The maximum mass comes out at  $M_{\rm max}= 2.02\, M_\odot$ with radius at $R = 11.86$ km. The central density is found to be  $n_{\rm cent} = 5.6\, n_0$. They are to be compared with the $V_{lowk}$ RG results of \cite{PKLMR} (for $n_{1/2}=2n_0$):  $M_{\rm max}=2.05 M_\odot$, $R = 12.19$ km, and $n_{\rm cent} = 5.1\, n_0$. There is very little difference -- and perhaps {\it no difference} within the approximations involved --  from the case of $n_{1/2}=2.0 n_0$ (and also that of $n_{1/2}=1.5 n_0$~\cite{dongetal}).
\begin{figure}[h]
\begin{center}
\includegraphics[width=4.3cm]{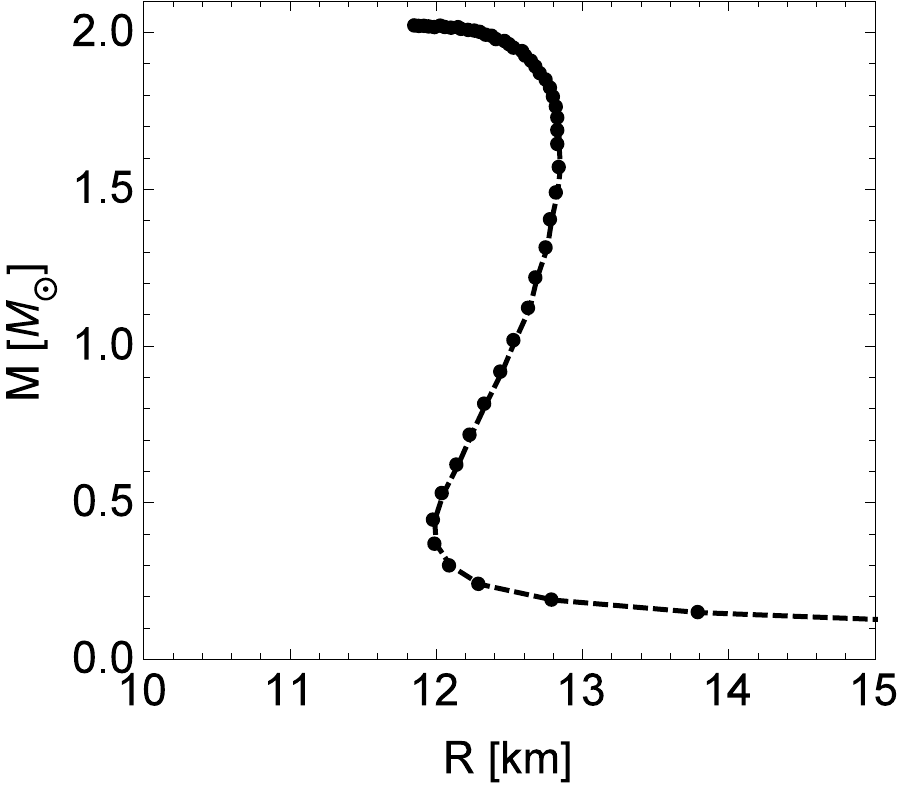}\includegraphics[width=4.2cm]{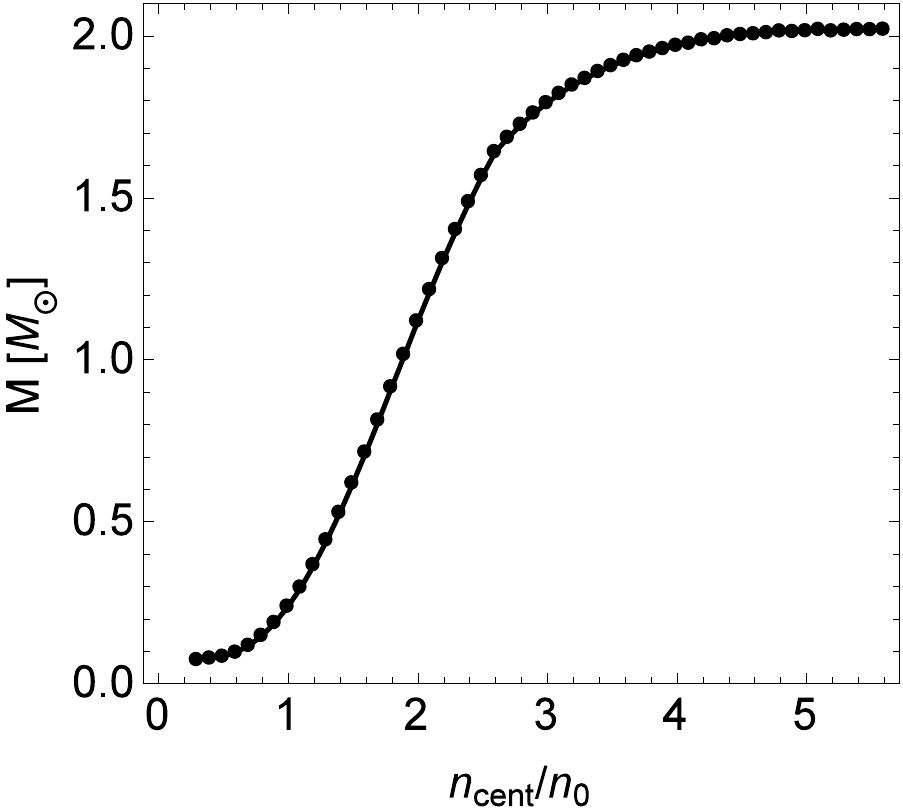}
\caption{ The mass vs. radius of the neutron star (left panel) in beta equilibrium and the mass vs. the central density of the neutron star (right panel).}\label{MvsR}
\end{center}
\end{figure}

These results are convincingly verifying that the pseudo-conformal EoS correctly captures the full $V_{lowk}$RG physics. They also
reinforce relative insensitivity of the standard star properties to the location of topology change $n_{1/2}$.

\sect{\bf Tidal deformability}
We now turn to the case of the tidal deformability.
 In a stark contrast,  the location of the topology change density  is found to strongly influence the tidal deformability in  the PCM.\footnote{This contrasts with other energy density functional models in which $\Lambda$ could be varied arbitrarily by fine-tuning parameters without affecting the basic structure of EoS.}

%
\begin{table}[h]
\begin{center}
\caption{Dimensional and dimensionless tidal  deformability:  $\lambda$ and $\Lambda$ }
\label{tidaldef}
\begin{center}
\begin{tabular}{c c c c c}
\hline
\hline
$M/M_{\odot}$ & $n_{cent}/n_0$   &  $\lambda/(10^{36}{\rm gcm^2s^2})$   & $\Lambda/100$  &   $ R/{\rm km}$ \\
\hline
1.12 & 2.0 & 4.16 & 22.5 & 12.64 \\
1.22 & 2.1 & 4.02 & 14.4 & 12.69 \\
1.31 & 2.2 & 3.86 & 9.45 & 12.76 \\
1.40 & 2.3 & 3.67 & 6.44 & 12.79 \\
1.49 & 2.4 & 3.46 & 4.52 & 12.83 \\
1.57 & 2.5 & 3.26 & 3.26 & 12.85 \\
\hline
\hline
\end{tabular}
\end{center}
\end{center}
\end{table}

 The results for the $\Lambda$ given by the PCM for $n_{1/2}=2.6 n_0$ are summarized  in Table.~\ref{tidaldef}  for the range of star masses involved.
Based on the effective mean-field cusp structure as predicted,  we expected $\Lambda$ to decrease,   when $n_{1/2}$ is increased to $2.6n_0$, from $\sim 800$ that was  found for $n_{1/2}=2.0 n_0$.  This expectation is confirmed: One indeed obtains a significant drop in $\Lambda$.  For $1.4\, M_\odot$, we find $\Lambda\sim 640$  with $n_{\rm cent} = 2.3\, n_0$. The radius for the $1.4 M_\odot$ star comes out to be $\sim 12.8$ km.


 It is noteworthy for assessing the principal mechanism in question  that the radius for the relevant star mass in the PCM is  insensitive to the location of $n_{1/2}$, hence to the cusp structure of the symmetry energy.\footnote{The radius is likely to depend on other details of the  star structure -- such as the crust etc. --  than the high density properties focused on in this paper. While the tidal deformability for the $1.4 M_\odot$ star is changed by roughly  25\% in going from $n_{1/2}=2n_0$ to $n_{1/2}=2.6n_0$, the radius is found to change by less than 1.5\%.}
 %

\sect{\bf Remarks}
 It is intriguing that the simple $E/A$, Eq.~(\ref{RII}), efficiently captures the complex structure of $\bf A$ \& $\bf B$ of $V_{lowk}$ RG in R-II. It seems remarkable also that  the appearance of the pseudo-conformal sound velocity $v_s^2\simeq 1/3$ at $n\gsim n_{1/2}$ as found in the $V_{lowk}$ RG calculation and confirmed by the pseudo-conformal model is  fully compatible with the observed properties of massive compact stars.   Given that converging to and staying at $v_s^2=1/3$ is impossible at non-asymptotic density unless there is a change of degrees of freedom~\cite{tews}, we are lead to that  the half-skyrmions  {\it must} play the role of non-hadronic degrees of freedom in the range of density in which strongly-coupled quarks, such as quarkyonic phase, are considered to figure~\cite{baymetal}.  The half-skyrmions in $n> n_{1/2}$ are confined by the magnetic monopole, an intrinsic property in QCD, hence a potential link to QCD degrees of freedom~\cite{cho}. This invites us to conjecture that the pseudo-conformal sound velocity is a signal in dense matter both for the emergence of scale symmetry and local flavor symmetry hidden in QCD~\cite{CCT,yamawaki2018} {\it and} for the manifestation of a Cheshire Cat phenomenon. This adds one more  to the growing evidence of Cheshire Cat phenomenon in dense hadronic matter, along with the superqualitons~\cite{qualiton},  vortices~\cite{alfordetal} etc. in the color-flavor-locked phase.

When the value of $\Lambda$ is measured  with a precision, it will provide  theoretically crucial information on the topology change density $n_{1/2}$, a.k.a, the hadron-quark crossover density.  At present there are no known clues, experimental or theoretical, as to how to locate it precisely. Should  the bound be tightened to a lower value than $\sim 600$ in the future, the PCM will provide a simple means to probe the range of $n_{1/2}$ compatible with the bound.  This will be the first impact of neutron-star physics on nuclear and particle physics in a quantity that cannot be accessed by reliable theories or experiments.

We are very grateful to Tom Kuo for suggesting the expression (\ref{RII}) and many useful  discussions. Y.L. Ma is supported in part by National Science Foundation of China (NSFC) under grant No. 1147571, 1147308 and Seeds Funding of Jilin University. Two of the authors, YLM and MR, would like to acknowledge the hospitality of APCTP, Korea where this paper was finalized.


\end{document}